\documentclass[aps,prd,preprint,groupedaddress]{revtex4-1}
\usepackage{graphicx}
\usepackage{epsfig}
\usepackage[caption=false]{subfig}

\begin{document}


\title{New observables for multiple-parton interactions measurements using Z + jets process at the LHC}


\author{M.~Bansal}%
\affiliation{D.A.V. College, Sector 10, Chandigarh, INDIA}

\author{S.~Bansal}
\email[]{sbansal@pu.ac.in}
\affiliation{Panjab University, Chandigarh, INDIA}

\author{R.~Kumar}
\email[]{raman\_phy@auts.ac.in}
\altaffiliation{Akal University, Talwandi Sabo, INDIA}
\affiliation{Panjab University, Chandigarh, INDIA}

\author{J. B.~Singh}
\affiliation{Panjab University, Chandigarh, INDIA}


\date{\today}

\begin{abstract}
Multiple-parton interactions play a vital role in hadron-hadron collisions. This paper presents a study 
of the multiple-parton interactions with  simulated Z + jets events in proton-proton collisions at a 
centre-of-mass energy of 13 TeV. The events are simulated with \textsc{powheg}~followed by hadronization 
and parton-showering using \textsc{pythia}8. The events with dimuon invariant mass in the range of 
60--120 GeV/$c^{\rm 2}$ are selected for the analysis. The charged particle jets, having minimum 
transverse momentum of 5 GeV/$c$ and absolute pseudo-rapidity less than 2, are used to construct 
the observables for measurements of the multiple-parton interactions. The proposed observables and 
phase-space region presented in this paper found to have enhanced sensitivity to multiple-parton 
interactions. The increased sensitivity to MPI will be lead to precise constraints on the parameters 
of the MPI models.
\end{abstract}


\maketitle

\section{Introduction}
The inelastic cross-section in hadron-hadron collisions, $e.g.$, at Large Hadron Collider (LHC), are dominated by 
semihard parton-parton scatterings producing particles with transverse momenta ($p_{\rm T}$) of few GeV/$c$.
At high energy collisions, the parton densities are large enough to cause a significant probability for two or more 
parton-parton scatterings within the same hadron-hadron collision.
The theoretical predictions for such multiple-parton interactions (MPI)~\cite{Sjostrand:1986ep} has been already 
confirmed by various experimental measurements at hadron colliders, $e.g.$, AFS, Tevatron and LHC. The study 
of MPI is important as it provides information on the parton-parton correlations and parton distributions in a 
hadron~\cite{Diehl:2011yj}. High $p_{\rm T}$/mass particles produced in the MPI also constitute as a background 
to new physics searches~\cite{Hussein:2006xr,SUSY}. The experimental identification of hadron and lepton, which 
are crucial to several measurements at LHC, is also affected by MPI.

MPI can not be completely described by perturbative quantum chromodynamics (QCD) and this requires a phenomenological 
description involving parameters that must be tuned with the help of data~\cite{Khachatryan:2015pea,Moraes:2007rq}. 
A wide range of experimental measurements is available for soft MPI ($p_{\rm T} \sim$ 0.5 GeV/$c$) as well as 
hard MPI of scale as high as the mass of W boson. For soft MPI 
analyses~\cite{uealice,ueatlas1,ueatlas2,ueatlas3,ueatlas4,uecms1,uecms2,uecms3,uecms4,uecdf,uedycms}, 
collective properties, $e.g.$, charged particle multiplicity and their scalar $p_{\rm T}$ sum are measured 
as a function of the event energy scale. These measurements are used to tune parameters of the MPI models. 
A better understanding of phenomenology of the MPI producing high $p_{\rm T}$ particles is equally important. 
It is needed to investigate how well the model parameters tuned using soft MPI measurements work with hard MPI.

A number of measurements has been performed for study of high-$p_{\rm T}$/mass particle production from 
MPI~\cite{jetua,jetafs,photon3jetd0,photon3jetcdf,4jetcdf,w2jetcms,w2jetatlas,photon3jetcms,wwcms},
$e.g.$, Z/W + jets, photon + jets, and 4-jets. These measurements explore the correlation observables, 
discussed in subsequent sections, between the particles produced from the first and second parton-parton 
scatterings. These observables show little sensitivity to MPI which lead to large systematic uncertainties
in the tuned model parameters~\cite{Khachatryan:2015pea}. In addition, inclusive MPI selection is required
for the theoretically correct estimation of the model parameters~\cite{Treleani,Seymour1,Seymour2}, but
experimental selection criteria restricts existing measurements to the double parton scattering (DPS) only.
This paper presents the study of Z + jets events to investigate new observables which consider higher order 
parton-parton interactions, in addition to DPS, and also have large sensitivity to the MPI. 

The outline of the paper is as follows. In Section~\ref{sec:EGnSC}, simulation of the Z + jets events and 
selection criteria is discussed. Section~\ref{sec:obs} describes the correlation observables used for the 
measurements of hard MPI. The results of the study of various observables for the measurements of the MPI 
are described in the Section~\ref{sec:results}. Section~\ref{sec:summary} summarizes the studies 
presented in this paper.

\section{Event generation and selection criteria}\label{sec:EGnSC} 

\subsection{Event generation}
Simulated events for Z + jets process, produced in pp collisions at 13 TeV, are generated up to NLO accuracy with 
\textsc{powheg}~\cite{Frixione:2007,PowhegW2J}. \textsc{powheg} uses the ``Multi-scale improved NLO'' (MiNLO) 
method~\cite{MINLO} which  describes well the jet productions associated with W/Z boson. These hard scattering events 
are hadronized and parton showered using \textsc{pythia}8~\cite{Sjostrand:2007gs}. The MPI are simulated with the
\textsc{pythia}8~\cite{Corke:2009pm}. In order to study the effect of MPI, Z + jets
events are also generated without the MPI simulation. The key features of the MPI model are:
\begin{itemize}
\item the ratio of the 2$\rightarrow$2 partonic cross section, integrated above a transverse momentum ($p_{\rm T}$) 
cutoff scale, to the total inelastic pp cross section, which is a measure of the amount of MPI. A factor with a free
parameter, ${{p_{\rm T}}_{\rm 0}}$, is introduced to regularize an otherwise divergent partonic cross section,

\begin{equation}\label{eq_P8v1}
\frac{\alpha_{\rm s}^{\rm 2}(p_{\rm T}^{\rm 2} + {{p_{\rm T}}_{\rm 0}}^{\rm 2})}{\alpha_{\rm s}^{\rm 2}(p_{\rm T}^{\rm 2})} \cdot \frac{p_{\rm T}^{\rm 4}}{(p_{\rm T}^{\rm 2} + {{p_{\rm T}}_{\rm 0}}^{\rm 2})^{\rm 2}}\ ,
\end{equation}

with

\begin{equation}\label{eq_P8v2}
{{p_{\rm T}}_{\rm 0}} (\sqrt{s}) = {{p_{\rm T}}_{\rm 0}} (\sqrt{s_{\rm 0}})
\left( \frac{\sqrt{s}}{\sqrt{s_{\rm 0}}} \right)^{\epsilon} \ .
\end{equation}

Here $\sqrt{s_{\rm 0}}$ = 1.8 TeV and `$\epsilon$' is a parameter characterizing the energy dependence of 
${{p_{\rm T}}_{\rm 0}}$.

\item The number of MPI in an event has a Poisson distribution with a mean that depends on the overlap of the matter
distribution of the hadrons in impact-parameter space. 
The present model uses the convolution of the matter distributions of the two incoming hadrons as 
an estimate of the impact parameter profile.
The overlap function is of the form e$^{-\rm bZ}$, where `b' is the impact parameter and `Z' is a free parameter.
\end{itemize}

The ATLAS A14 tune~\cite{ATLAS:2012uec} with NNPDF2.3LO set of parton distribution functions (PDF) is used.
This set of MPI model parameters is obtained by fitting the underlying event data obtained at the LHC.
There are other MPI models for \textsc{pythia}8 and various Monte-Carlo event generators also exist with MPI model different 
than \textsc{pythia}8, $i.e.$, \textsc{sherpa}~\cite{Gleisberg:2008ta}, \textsc{herwig}++~\cite{Bahr:2008pv} , $etc$.
But, the methodology for MPI measurement, presented in this paper, is not specific to choice of MPI model or Monte-Carlo event 
generators. This methodology is devised as per experimental feasibility of the measurement which depends upon the control 
of the single parton scatterings (SPS) background. Also the correlation between SPS background and boost of Z- boson is 
independent of MPI modeling. Therefore, conclusion of the presented paper is independent of the choice of MPI model.


\subsection{Event selection}
Events with well identified Z-candidates are selected using the kinematic properties of the muons produced 
by the decay of the Z-boson. The kinematic selection criteria used for Z-candidate events is:
\begin{itemize}
 \item
   exactly two muons with $p_{\rm T}$ larger than 20 GeV/$c$ and absolute pseudo-rapidity ($\eta$) less than 2.5.
\item
   Invariant mass of the two muons is defined as:
    \begin{equation}
     M_{\mu\mu} = \sqrt{ (E^1+E^2)^2 - (p_x^1+p_x^2)^2 - (p_y^1+p_y^2)^2 - (p_z^1+p_z^2)^2},
    \end{equation}
    where ($E^1$,$p_x^1$,$p_y^1$,$p_z^1$) and ($E^2$,$p_x^2$,$p_y^2$,$p_z^2$) are the four-momenta of the two muons.
    The events with invariant mass around Z-mass within the window of 60-120 GeV/${c^{\rm 2}}$ are selected.
\end{itemize}
This kinematic selection criteria is motivated from the trigger and the background constraints at the LHC experiments.

Jets are clustered using anti-$k_T$ algorithm~\cite{Cacciari:2008gp} with the radius parameter equal to 0.5. 
Only charged particles, with $|\eta|<$ 2, are used for the jet clustering because the charged particles are 
reconstructed in the tracking detector and hence identification efficiency is large even at 
very low $p_{\rm T}$~\cite{uedycms}. A condition of $|\eta|<$ 2 is used considering the tracker acceptance 
of the LHC experiments. The cluster jets are required to have minimal $p_{\rm T}$ of 20 GeV/$c$ and $\eta<$ 2. 
Jets with lower $p_{\rm T}$ threshold are also considered to study the dependency on energy scale of the MPI.

\section{Observables}\label{sec:obs}
For soft MPI measurements, inclusive properties, $e.g.$, charged particle multiplicity and scalar sum of 
$p_{\rm T}$ of charged particles as a function of $p_{\rm T}$/mass of the leading object, are used. 
These distributions are fitted using \textsc{professor}~\cite{Buckley:2009bj} to estimate the tuned 
parameters of MPI models. The hard MPI analyses are restricted to DPS only and use correlation observables 
between decay products of the first and second hard interactions. These observables are based on the assumption
that MPI are independent of each other. The hard MPI measurements at Tevatron~\cite{photon3jetd0,photon3jetcdf}
and LHC~\cite{w2jetcms,w2jetatlas,photon3jetcms} are performed using various correlation observables, 
$e.g.,\Delta$S and $\Delta p_{\rm T}^{\rm rel}$, which can be defined for Z + jets process as follows:

\begin{itemize}
 \item
 the relative $p_{\rm T}$-balance ($\Delta p_{\rm T}^{\rm rel}$) between two jets,
 is defined as: 
 \begin{equation}
  \Delta p_{\rm T}^{\rm rel} = \frac{|\vec{p_{\rm T}}{\rm (j1)} + \vec{p_{\rm T}}{\rm (j2)}|}{|\vec{p_{\rm T}}{\rm (j1)}| + |\vec{p_{\rm T}}{\rm (j2)}|},
 \end{equation}
 where $\vec{p_{\rm T}}{\rm (j1)}$ and $\vec{p_{\rm T}}{\rm (j2)}$ represent transverse momentum vectors of
 leading and subleading jets respectively. 
 
 \item 
 Azimuthal separation between Z boson and di-jet ($\Delta$S) is defined as:
 \begin{equation}
  \Delta {\rm S} = \cos^{\rm -1}(\frac{\vec{p_{\rm T}}{\rm (Z)}\cdot \vec{p_{\rm T}}{\rm (di-jet)}}{|\vec{p_{\rm T}}{\rm (Z)}|~|\vec{p_{\rm T}}{\rm (di-jet)}|}),
 \end{equation}
 
 where $\vec{p_{\rm T}}{\rm (Z)}$ and $\vec{p_{\rm T}}{\rm (di-jet)}$ represent transverse momentum vectors of the 
 Z boson and di-jet system respectively. 
\end{itemize}

For Z + jets events (jets are required to have $p_{\rm T}$ larger than 20 GeV/$c$ and $|\eta|<$ 2),  the 
distributions of $\Delta p_{\rm T}^{\rm rel}$ and $\Delta$S observables are shown in Fig.~\ref{corr_gen}(a)
and Fig.~\ref{corr_gen}(b) respectively. In DPS events, at leading order (LO),  the two jets balance each other
 and $\Delta p_{\rm T}^{\rm rel}$ is small, which is not the case for events from SPS.
 In DPS events, transverse momentum vectors of the two objects are randomly 
 oriented. It leads to a flat distribution of $\Delta$S observable for DPS events. For SPS events, the Z boson
 and di-jet are supposed to balance each other at LO and hence $\Delta$S observable is expected to have values 
 closer to $\pi$ radians. 
  MPI contribution is extracted by a template fit method~\cite{w2jetcms,w2jetatlas}
 as well as by fitting~\cite{Khachatryan:2015pea} the distributions with \textsc{professor}.
In this paper, jet multiplicity observable is investigated for the MPI measurements. Jet multiplicity does not 
require any exclusive selection on MPI and sensitivity is also better, as will be discussed 
Section~\ref{sec:results}, compared to the correlation observables.

\section{Results and discussions}\label{sec:results}
\begin{figure}[htbp]
\begin{center}
\subfloat[]{\includegraphics[width=0.5\textwidth]{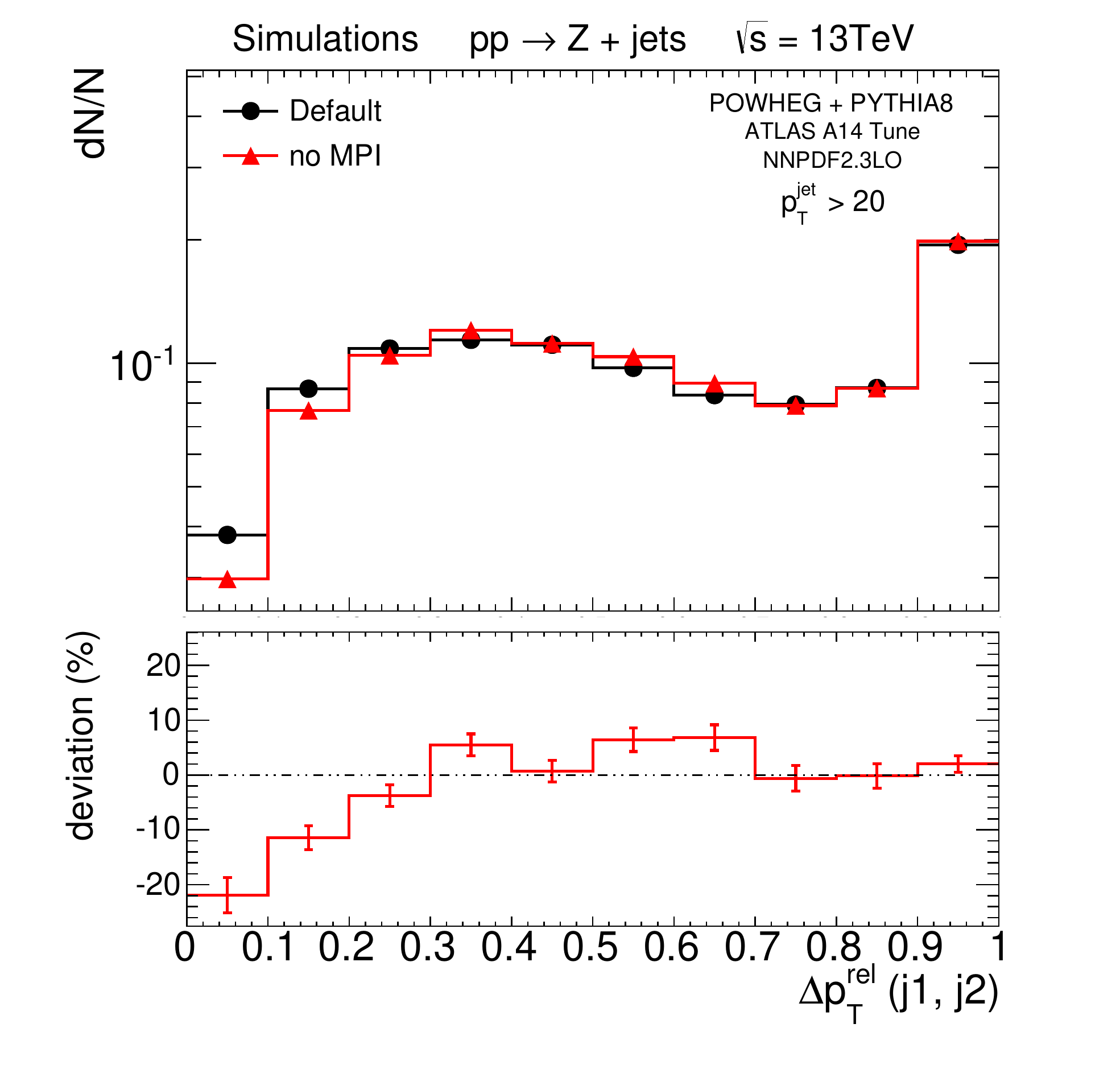}}
\subfloat[]{\includegraphics[width=0.5\textwidth]{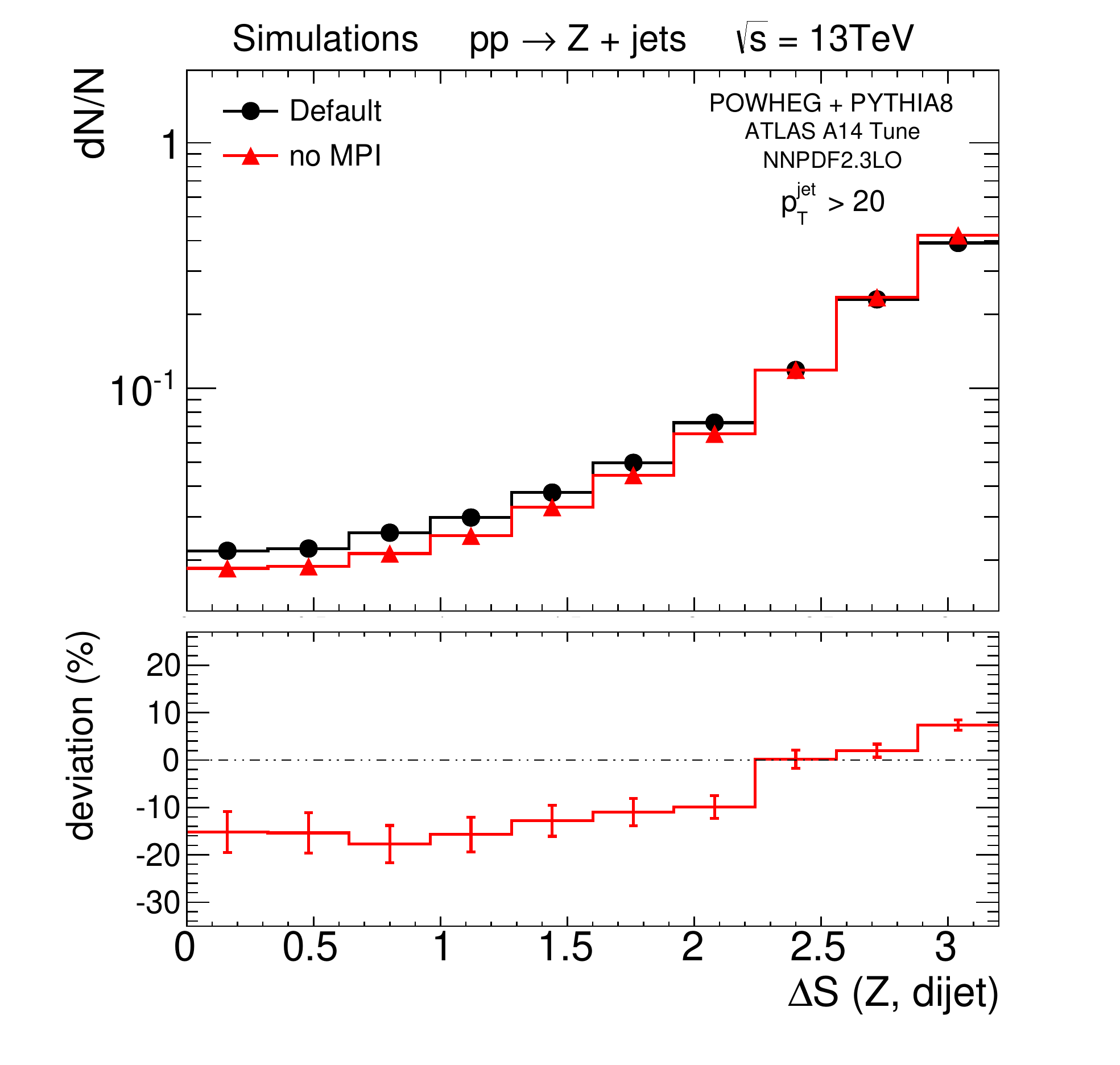}}
\end{center}
\caption {{The distributions of correlation observables, (a) $\Delta p_{\rm T}^{\rm rel}$ and (b) $\Delta$S are compared for 
events with and without MPI. The events from Z + jets processes are generated using \textsc{powheg}, parton showered and 
hadronized with \textsc{pythia}8. Jets with $p_{\rm T}$ larger than 20 GeV/$c$ are considered. The ratio plot in the bottom 
panel shows the deviations of the distributions after switching off MPI.}} \label{corr_gen}
\end{figure}

\begin{figure}[htbp]                                                                  
\begin{center}
\subfloat[]{\includegraphics[width=0.5\textwidth]{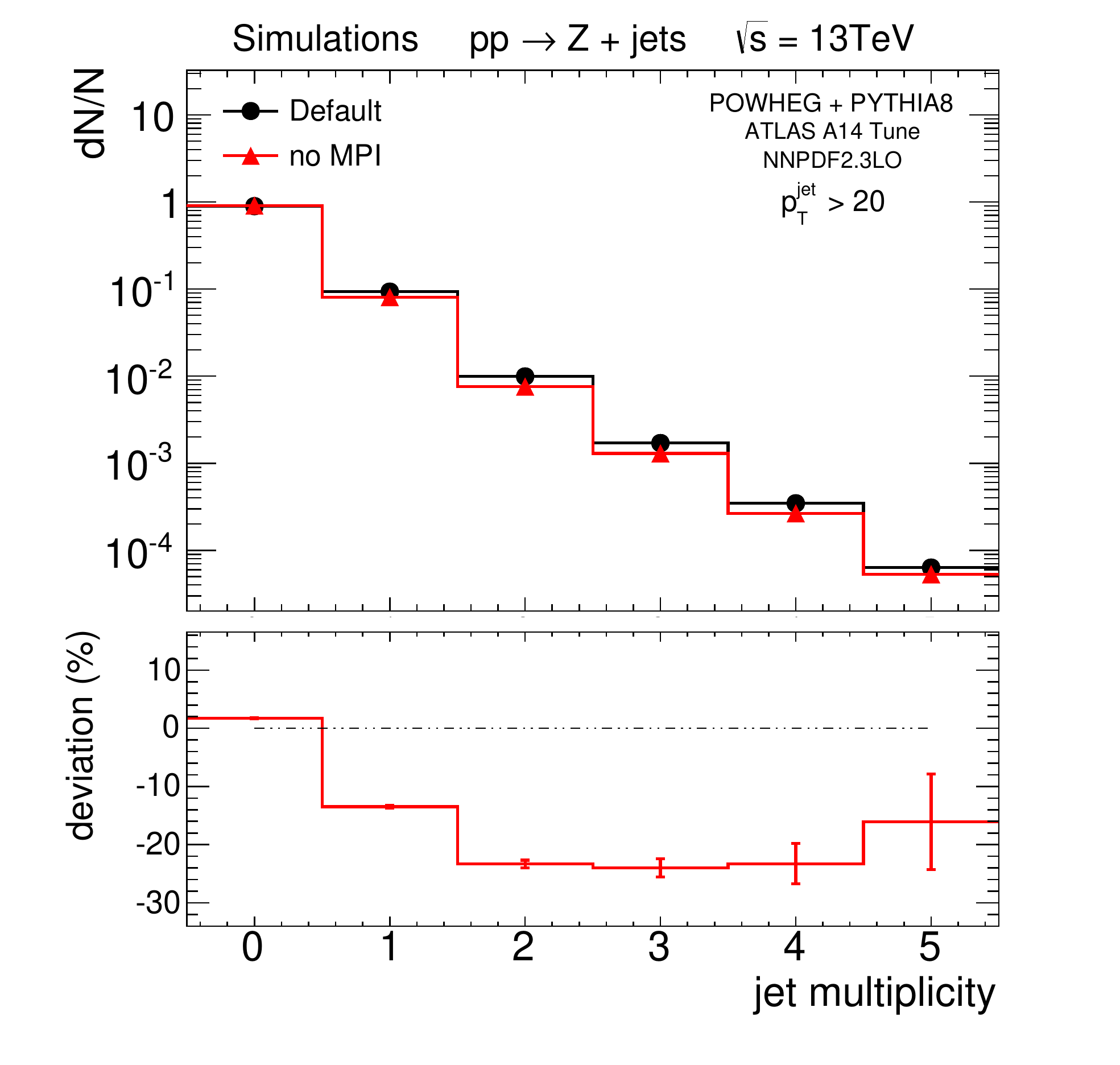}}
\subfloat[]{\includegraphics[width=0.5\textwidth]{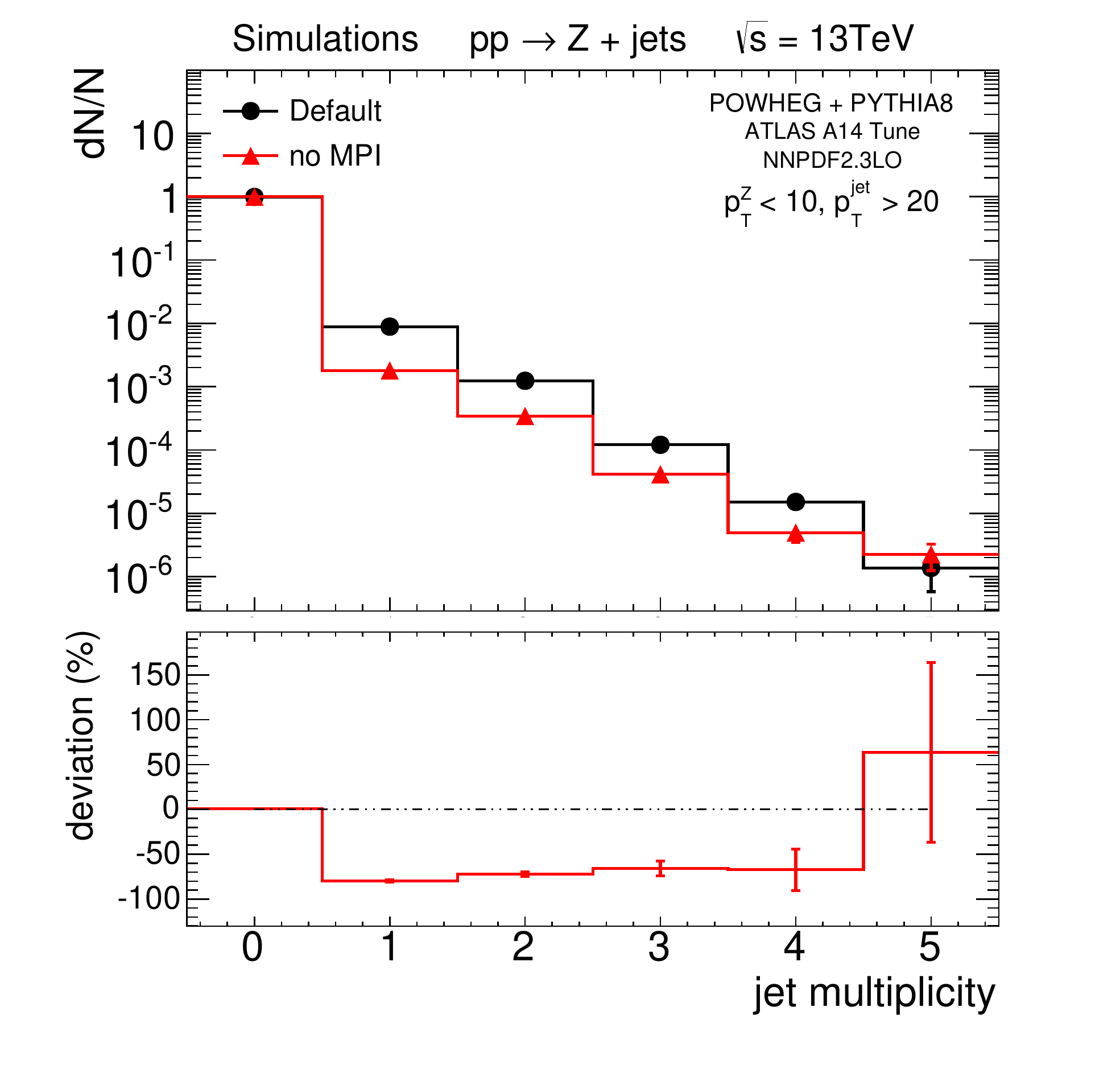}}
\caption {{Jet multiplicity distributions are compared for events with and without MPI. The events from Z + jets 
processes are generated using \textsc{powheg}, parton showered and hadronized with \textsc{pythia}8. Jets with 
$p_{\rm T}$ larger than 20 GeV/$c$ are considered. The 
distributions are shown (a) without any condition on $p_{\rm T}^{\rm Z}$ and (b) with $p_{\rm T}^{\rm Z}$ 
less than 10 GeV/$c$. The ratio plot in the bottom panel shows deviations of the distributions after switching off MPI.}} \label{jm_def}
\end{center}
\end{figure}

\begin{figure}[htbp]                                                                  
\begin{center}
\subfloat[]{\includegraphics[width=0.5\textwidth]{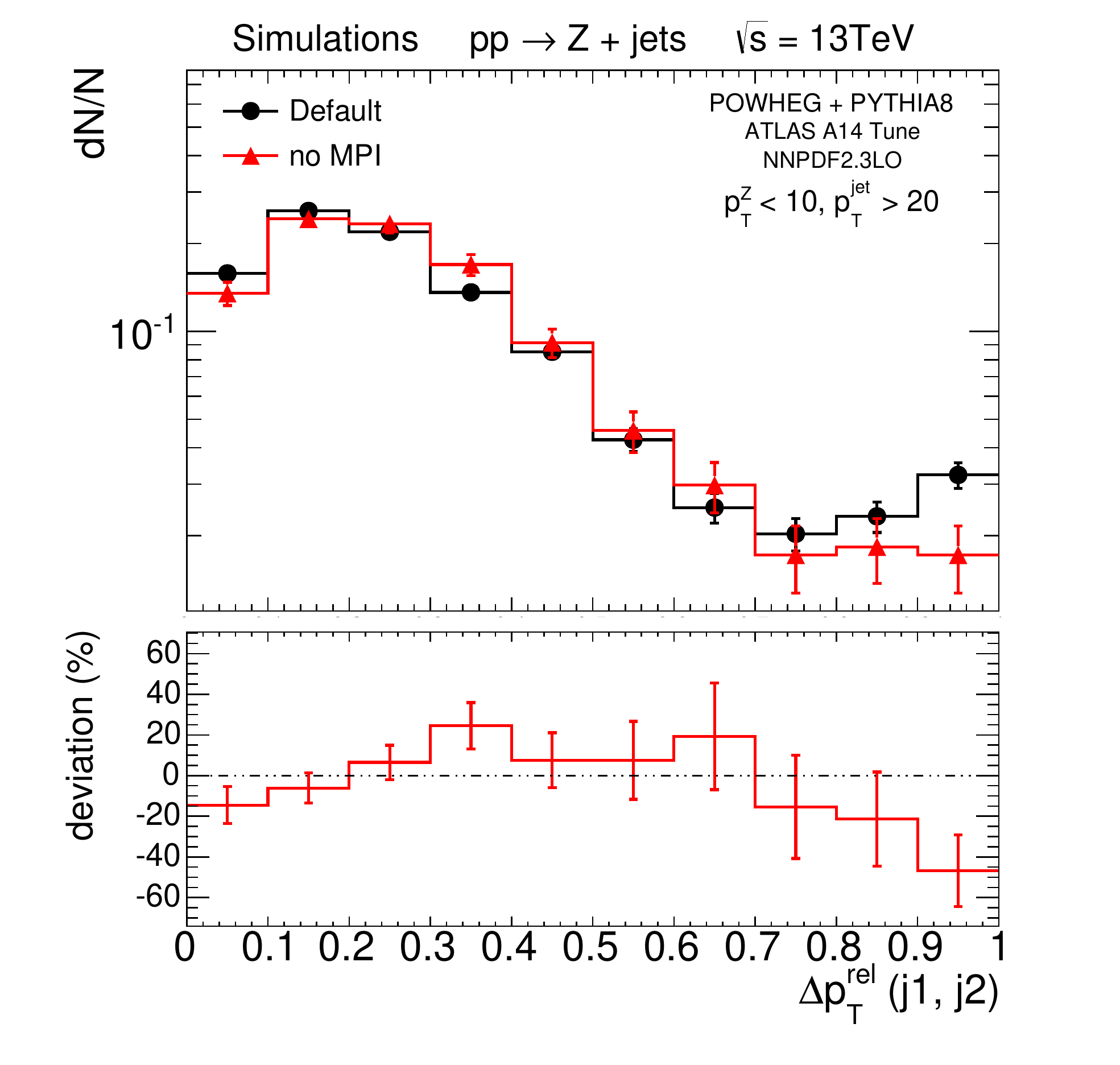}}
\subfloat[]{\includegraphics[width=0.5\textwidth]{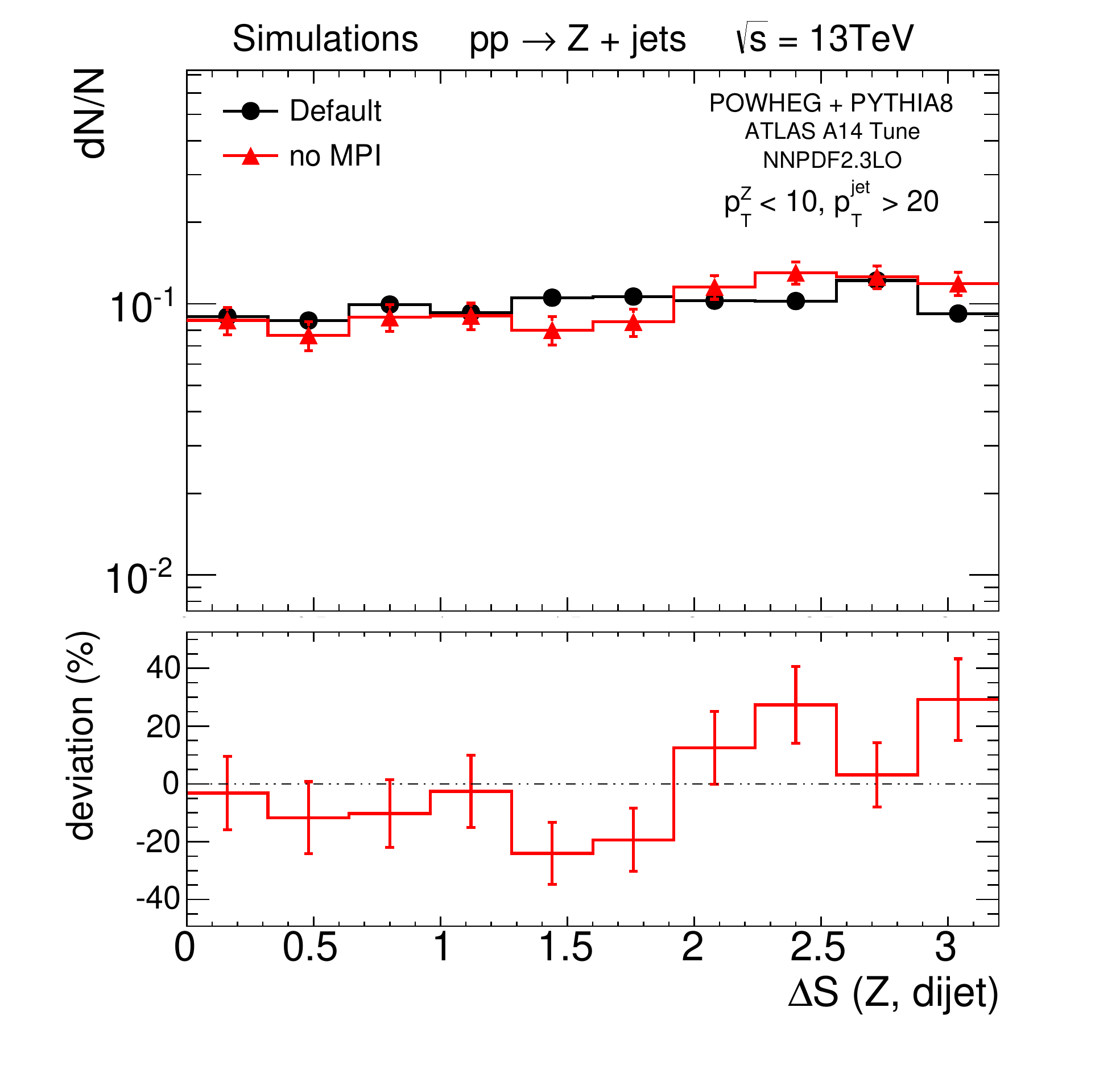}}
\caption {{The distributions of correlation observables, (a) $\Delta p_{\rm T}^{\rm rel}$ and (b) $\Delta$S are compared for events 
with and without MPI. The events from Z + jets processes are generated using \textsc{powheg}, parton showered and hadronized with 
\textsc{pythia}8. Jets with $p_{\rm T}$ greater than 20 GeV/$c$ are considered and $p_{\rm T}^{\rm Z}$ is required 
to be less than 10 GeV/$c$. The ratio plot in the bottom panel shows deviations of the distributions after switching off MPI.}} \label{corr_Zpt10}
\end{center}
\end{figure}

Figure~\ref{corr_gen}(a) and Fig.~\ref{corr_gen}(b) shows the distributions of $\Delta p_{\rm T}^{\rm rel}$ 
and $\Delta$S observables for 
Z + jets events, with the selection criteria as mentioned in Section~\ref{sec:EGnSC}, with and without MPI simulation. 
A visible effect ($\sim$20\%) of MPI is observed at the lower values of $\Delta p_{\rm T}^{\rm rel}$ 
and up to 15\% for $\Delta$S. MPI also affect the jet multiplicity distribution up to 25\% as shown in 
Fig.~\ref{jm_def}~(a). There is no apparent gain in MPI sensitivity by using jet multiplicity, but it will  
lead to inclusive MPI selection unlike the correlation observables. The sensitivity to MPI is low due to large background from the SPS.  
For SPS events, there is a direct correlation between $p_{\rm T}$ of Z-boson ($p_{\rm T}^{\rm Z}$) and associated jets. 
Therefore, SPS background can be reduced by applying an upper cut on the $p_{\rm T}^{\rm Z}$. 
A gain, in terms of MPI sensitivity by a factor 3--4, is achieved by
applying an upper cut of 10 GeV/$c$ on the $p_{\rm T}^{\rm Z}$ as depicted in Fig.~\ref{jm_def} (b). 
The correlation observables are also expected to show the same gain with upper cut 
on p$_{\rm T}^{\rm Z}$, but it is observed that sensitivity decreases for the correlation observables.
This is because of the fact that kinematics of SPS events, which survived after an upper on $p_{\rm T}^{\rm Z}$, 
are more like MPI events. 
There will a good fraction of events having two jets which balance each other but not the Z-boson.
These two jets can be randomly oriented with respect to Z-direction. 
Thus, for the selected SPS events correlation observables will have same distributions as for MPI events.
This observation is visible from Fig.~\ref{corr_Zpt10}(a) and Fig.~\ref{corr_Zpt10}(b) which shows the 
comparison of correlation observables in Z + jets with and without MPI after requiring $p_{\rm T}^{\rm Z}<$ 10 GeV/$c$. 

To investigate further if jet multiplicity distribution will give better estimation of individual MPI model parameters, 
following parameters are studied:

\begin{itemize}

 \item {${{p_{\rm T}}_{\rm 0}} (\sqrt{s_{\rm 0}})$: value of $p_{\rm T}$ cut-off scale at 
 reference energy scale, $\sqrt{s_{\rm 0}}$= 1.8 TeV. The ATLAS A14 tune with NNPDF2.3LO PDF set
 uses 2.09 GeV/$c$ as its default value.}

 \item {Z: free parameter for the impact parameter profile of the colliding hadron beams as discussed in Section~\ref{sec:EGnSC}.
The default value for `Z' is 1.85 cm$^{\rm -1}$.}

 \item {$\alpha_{\rm S}^{\rm MPI}$: value of strong coupling constant for MPI chosen at the mass of Z-boson. The default 
 value of this free parameter is 0.126.}

\end{itemize}

\begin{figure}[htbp]                                                                  
\begin{center}
\subfloat[]{\includegraphics[width=0.49\textwidth]{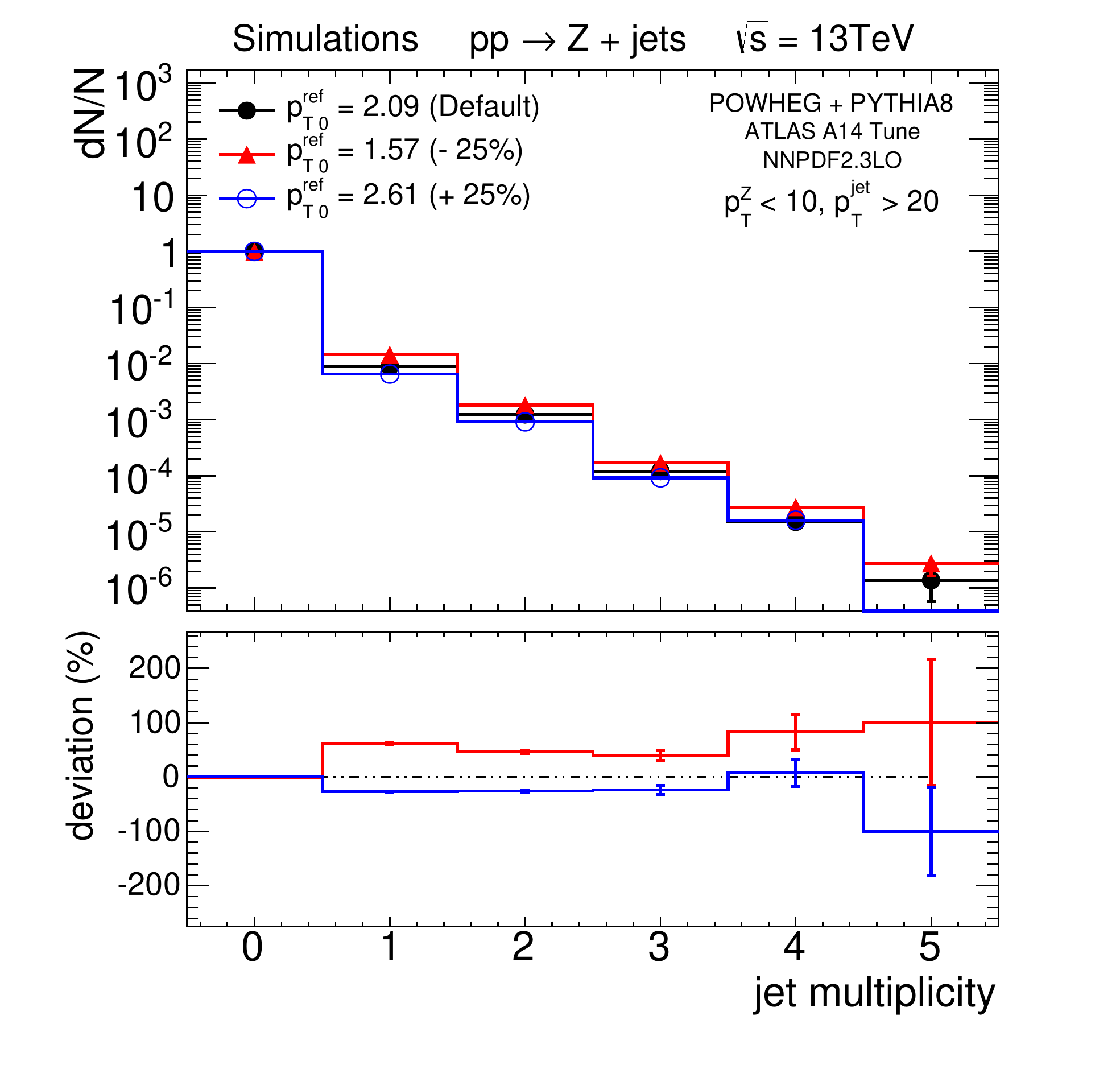}}
\subfloat[]{\includegraphics[width=0.49\textwidth]{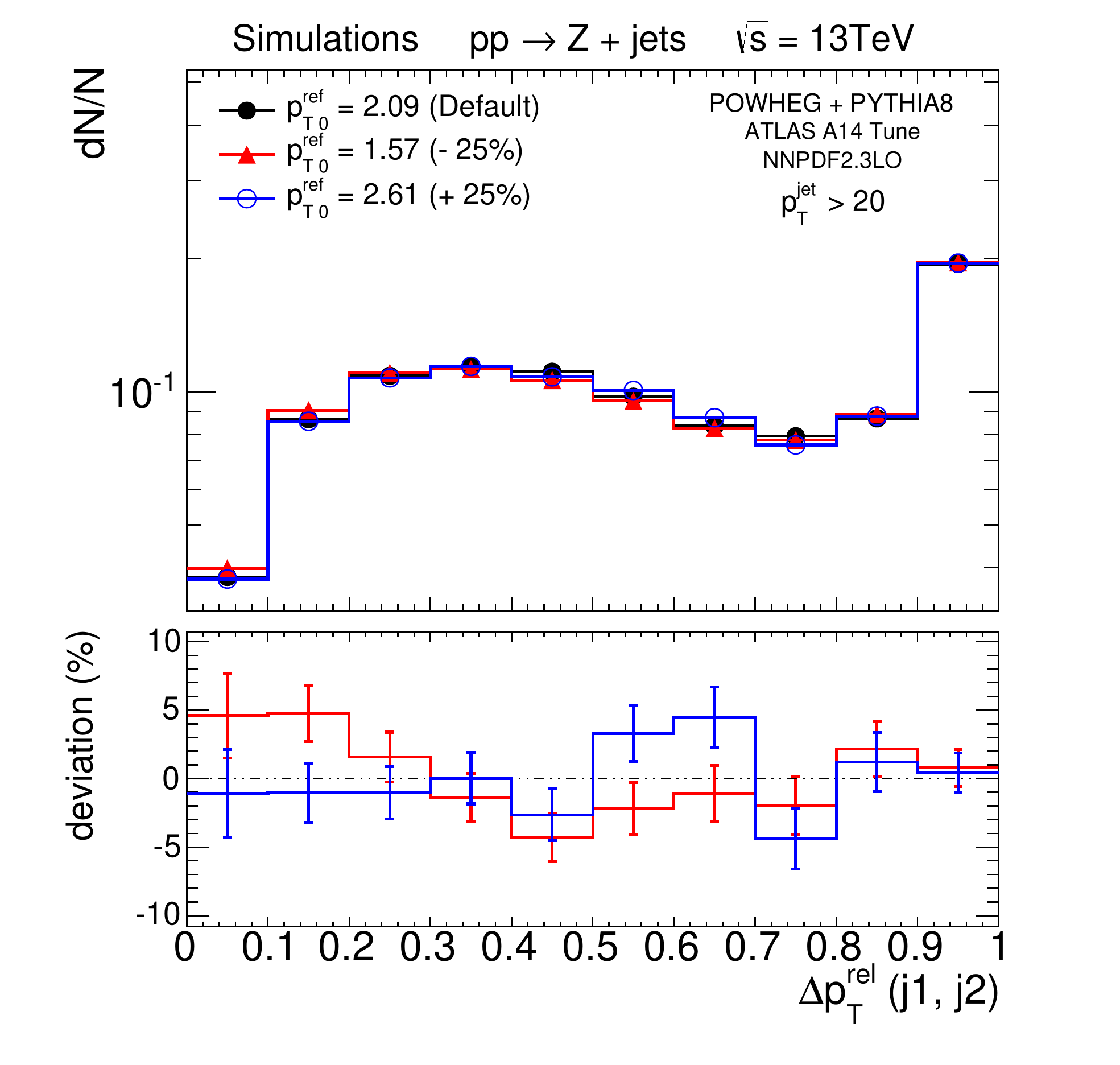}}\\
\subfloat[]{\includegraphics[width=0.49\textwidth]{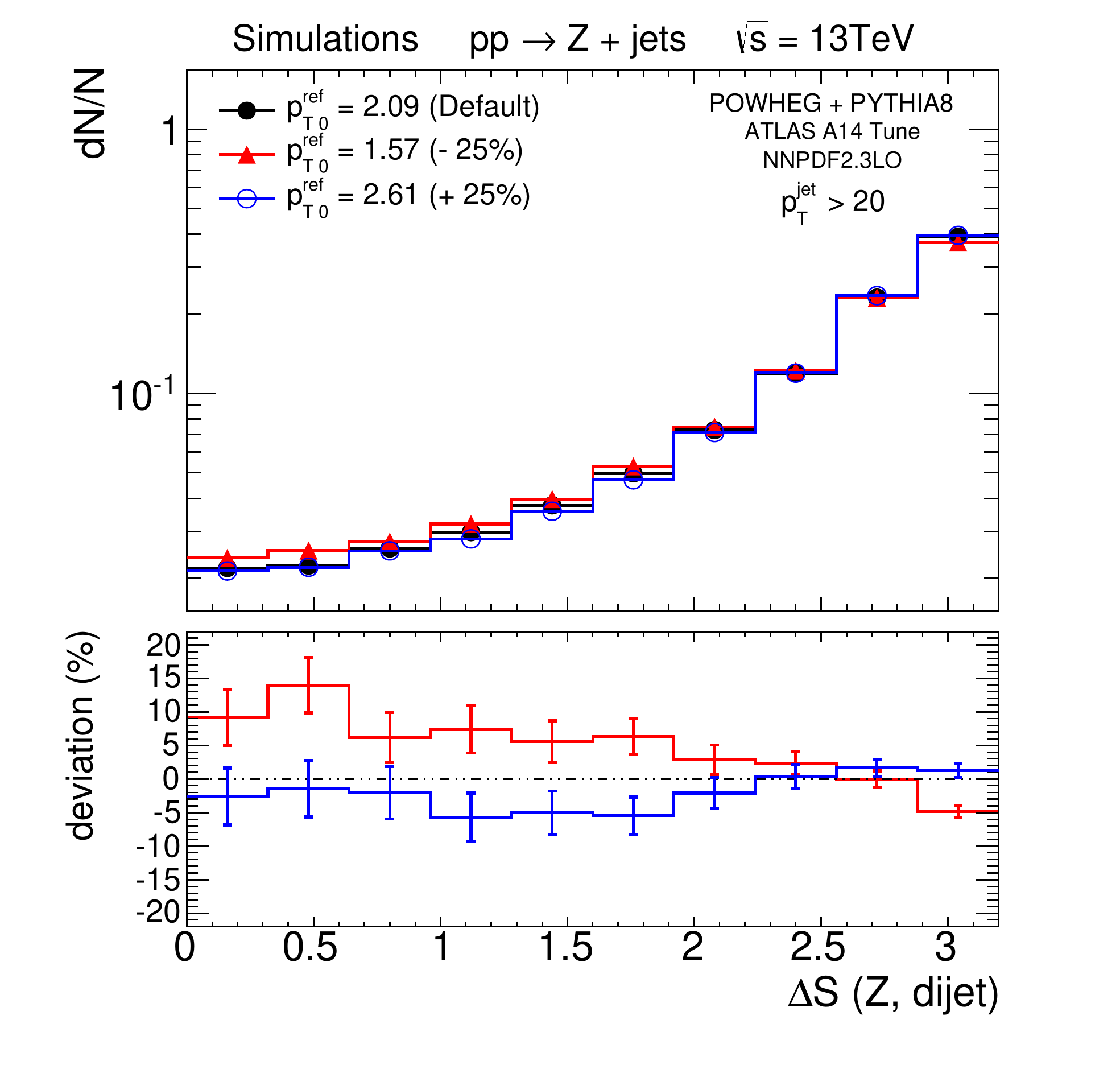}}
\caption {{The distributions of (a) jet multiplicity, (b) $\Delta p_{\rm T}^{\rm rel}$ and (c) $\Delta$S observables 
are compared with variation ($\pm$25\%) in 
${{p_{\rm T}}_{\rm 0}} (\sqrt{s_{\rm 0}})$. The events from Z + jets processes are generated using \textsc{powheg}5, 
parton showered and hadronized with \textsc{pythia}8. The value of $p_{\rm T}^{\rm Z}$ is required to be less than 
10 GeV/$c$ and jets with $p_{\rm T}$ larger than 20 GeV/$c$ are considered. The ratio plot in the bottom panel shows 
deviations of the distributions after switching off MPI.  }} \label{jm_Avar}
 \end{center}
\end{figure}

\begin{figure}[htbp]                                                                  
\begin{center}
\subfloat[]{\includegraphics[width=0.49\textwidth]{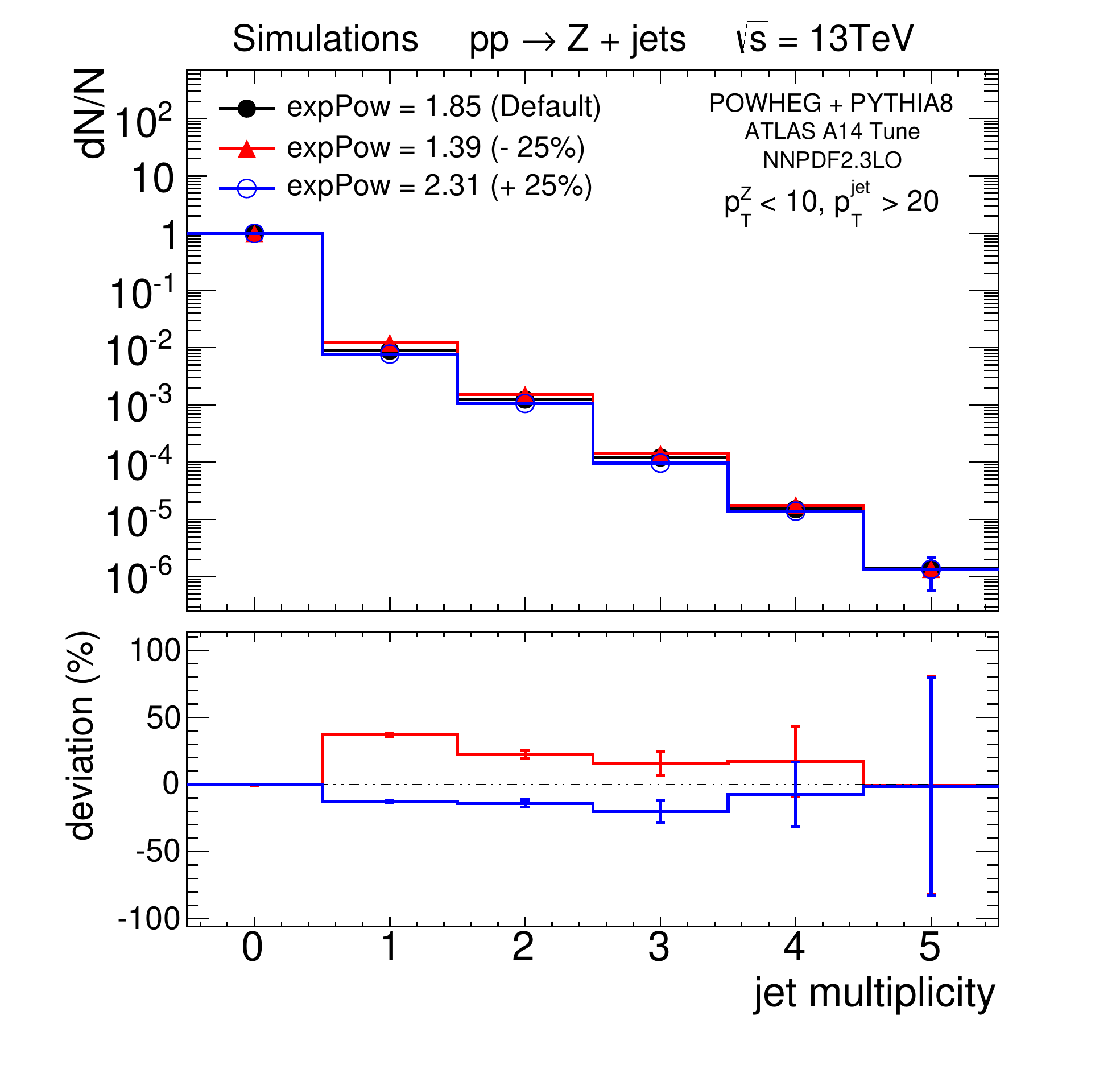}}
\subfloat[]{\includegraphics[width=0.49\textwidth]{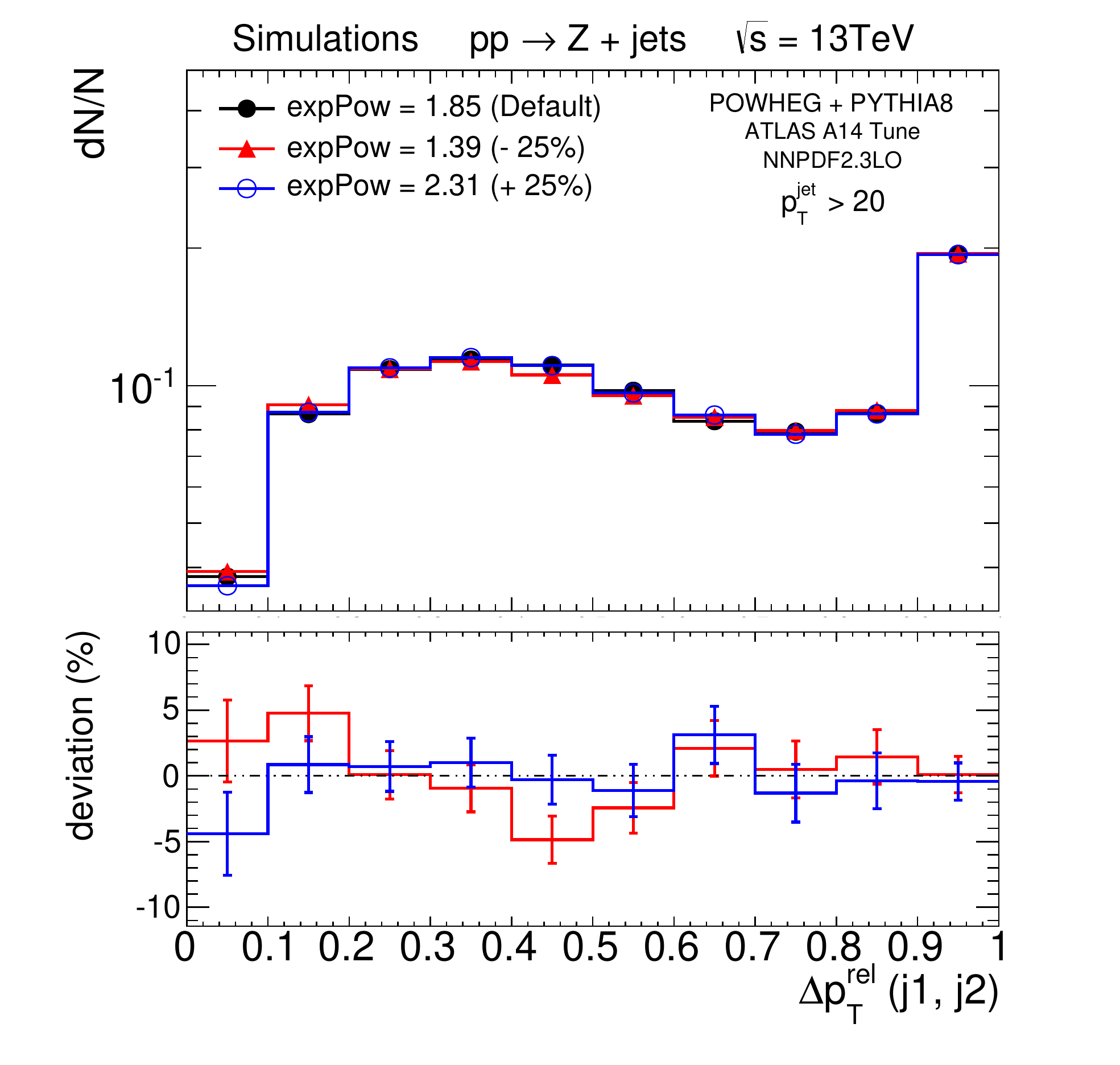}}\\
\subfloat[]{\includegraphics[width=0.49\textwidth]{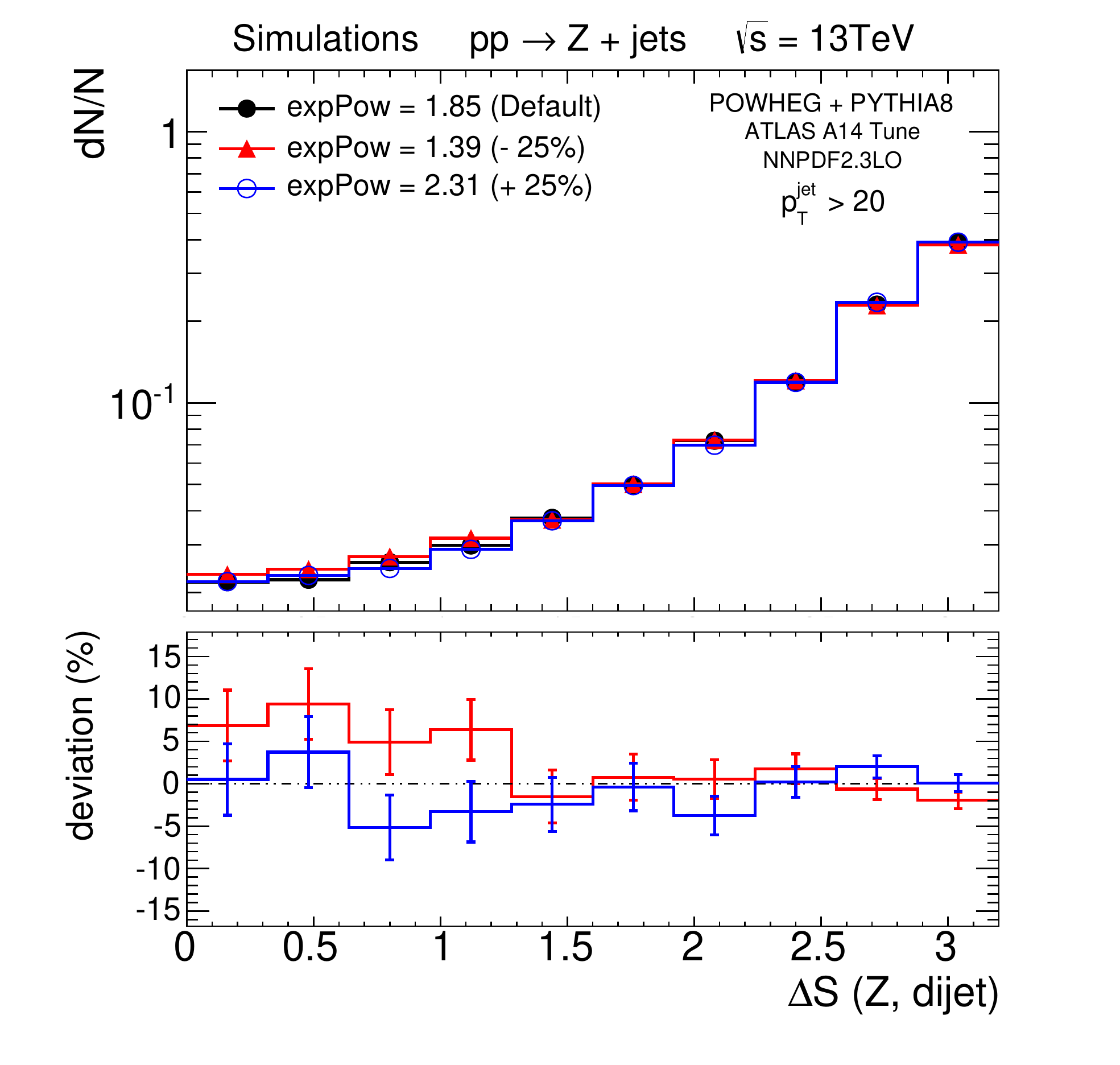}}
\caption {{The distributions of (a) jet multiplicity, (b) $\Delta p_{\rm T}^{\rm rel}$ and (c) $\Delta$S observables are 
compared with variation ($\pm$25\%) in 
`Z'. The events from Z + jets processes are generated using \textsc{powheg}, parton showered and 
hadronized with \textsc{pythia}8. The value of $p_{\rm T}^{\rm Z}$ is required to be less than 10 GeV/$c$ 
and jets with $p_{\rm T}$ larger than 20 GeV/$c$ are considered. The ratio plot in the bottom panel shows 
deviations of the distributions after switching off MPI. }} \label{jm_Cvar}
 \end{center}
\end{figure}

\begin{figure}[htbp]                                                                  
\begin{center}
\subfloat[]{\includegraphics[width=0.49\textwidth]{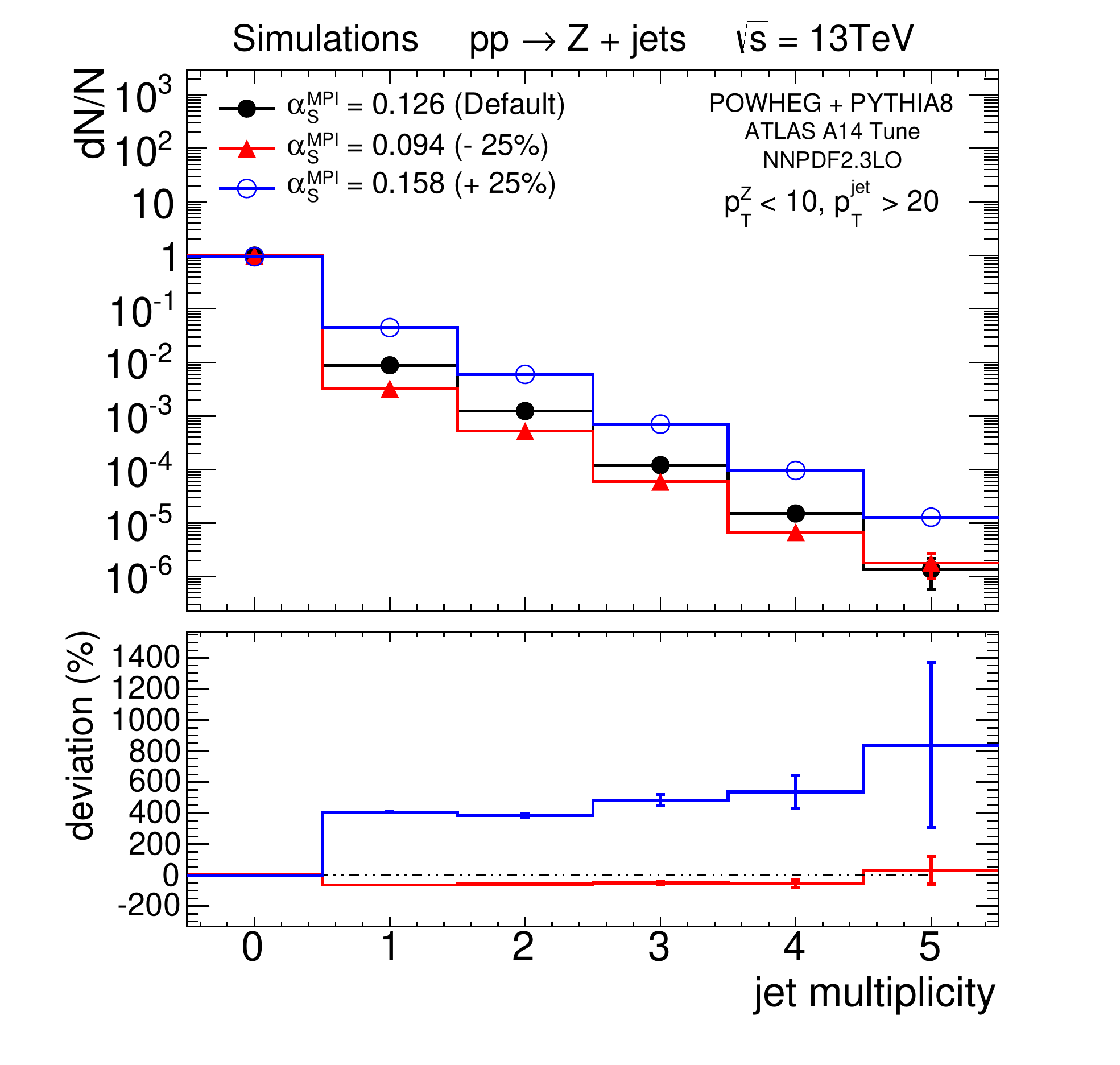}}
\subfloat[]{\includegraphics[width=0.49\textwidth]{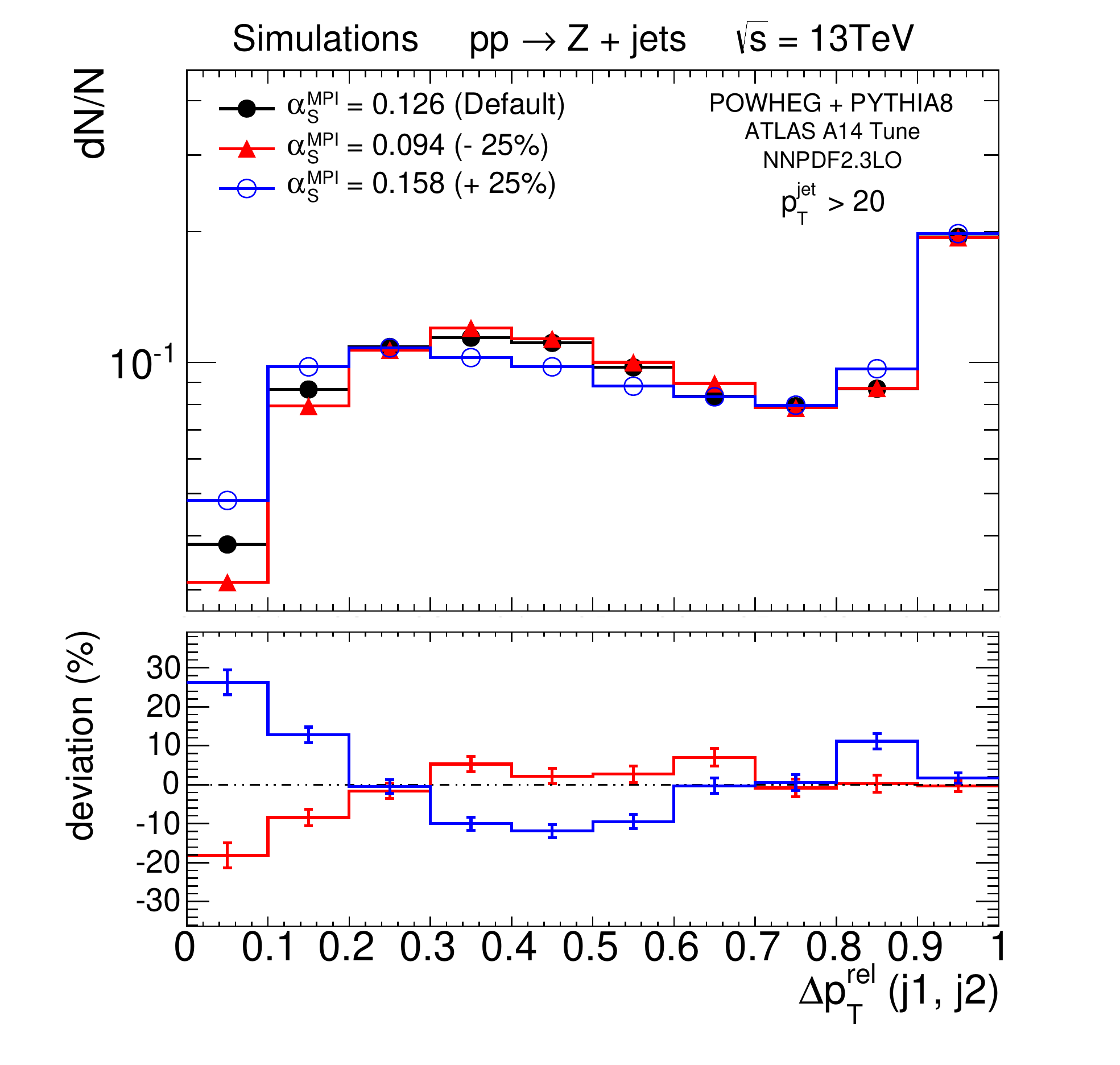}}\\
\subfloat[]{\includegraphics[width=0.49\textwidth]{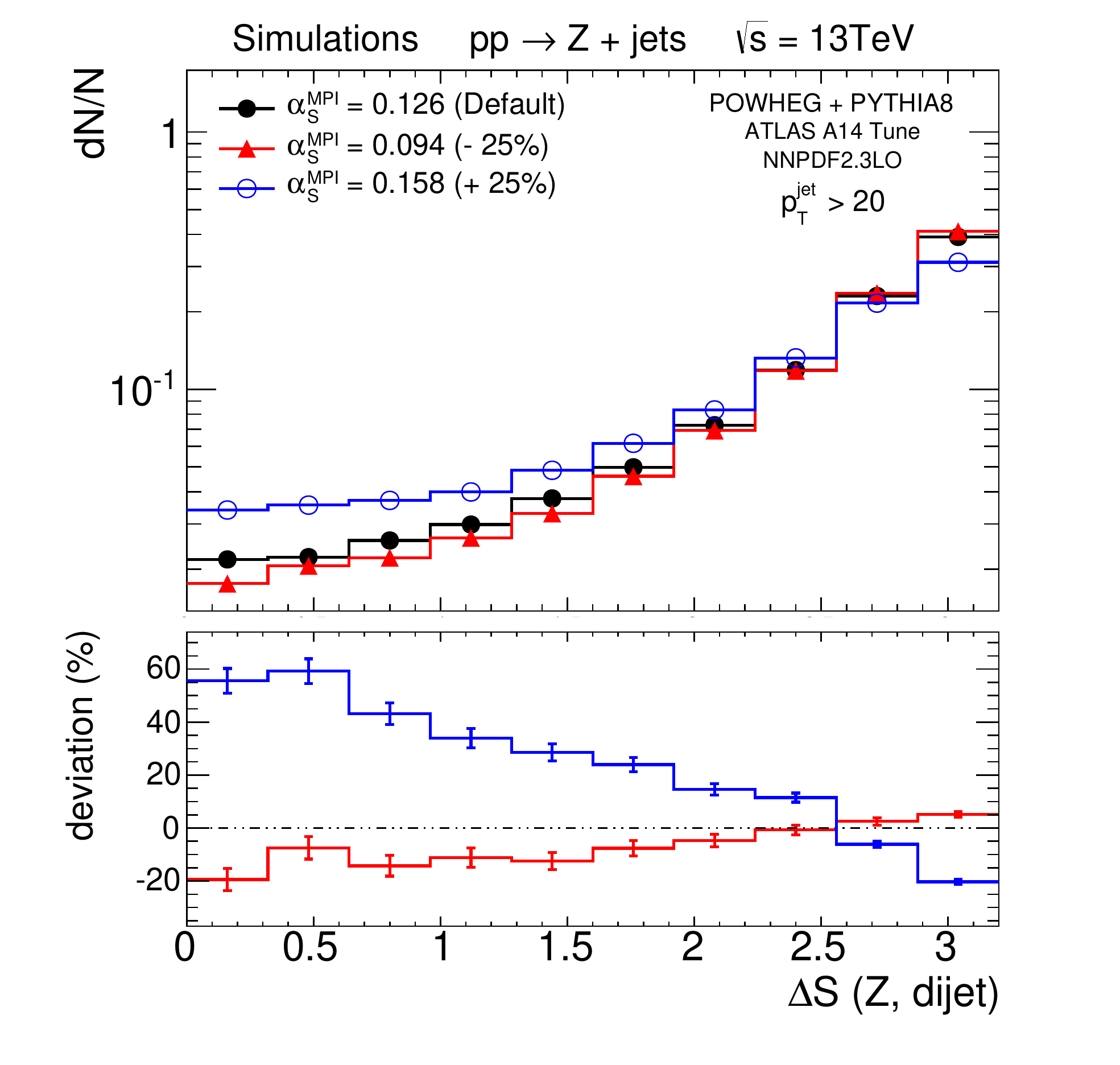}}
\caption {{The distributions of (a) jet multiplicity, (b) $\Delta p_{\rm T}^{\rm rel}$ and (c) $\Delta$S observables are 
compared with variation ($\pm$25\%) in 
$\alpha_{\rm S}^{\rm MPI}$. The events from Z + jets processes are generated using \textsc{powheg}, parton showered and 
hadronized with \textsc{pythia}8.  The value of $p_{\rm T}^{\rm Z}$ is required to be less than 10 GeV/$c$ and 
jets with $p_{\rm T}$ larger than 20 GeV/$c$ are considered. The ratio plot in the bottom panel shows 
deviations of the distributions after switching off MPI. }} \label{jm_Dvar}
 \end{center}
\end{figure}

These parameters are varied independently by $\pm$25$\%$ to study their effect on sensitivity of the observables.
The values of these parameters along with their default values are given in Table~\ref{tab_varParam}.
The effect of variation in ${{p_{\rm T}}_{\rm 0}} (\sqrt{s_{\rm 0}})$ on sensitivity of jet multiplicity, 
$\Delta p_{\rm T}^{\rm rel}$ and $\Delta$S observables is shown in Fig.~\ref{jm_Avar}(a), Fig.~\ref{jm_Avar}(b)
and Fig.~\ref{jm_Avar}(c) respectively. 
The correlation observables show the deviations up to 5--10\%, whereas jet multiplicity shows the 
deviations up to 60\%. Figure~\ref{jm_Cvar}(a), Fig.~\ref{jm_Cvar}(b) and Fig.~\ref{jm_Cvar}(c) shows the effect 
of MPI parameter `Z' on the  jet multiplicity, $\Delta p_{\rm T}^{\rm rel}$ and $\Delta$S observables respectively.
The jet multiplicity distribution shows the maximal variation of 40\% against the little effect  ($<$10 \%) on the correlation observables.
Similarly, with the variation of $\alpha_{\rm S}^{\rm MPI}$, it is observed that correlation
observables show sensitivity as high as 60\% (Fig.~\ref{jm_Dvar}(b) and Fig.~\ref{jm_Dvar}(c)), whereas jet multiplicity shows 
sensitivity even larger than 300\%, as shown in Fig.~\ref{jm_Dvar}~(a). 
It is clear that same variation in MPI model parameters gives large affect on the jet multiplicity as compared to the correlation observables.
As discussed earlier, enhanced MPI sensitivity for jet multiplicity distribution is due to the suppression of the SPS background with an upper cut on the $p_{\rm T}^{\rm Z}$.
This large sensitivity to the MPI parameters will lead to more accurate estimation of the model parameters with reduced systematic uncertainties. 

\begin{figure}[htbp]                                                                  
\begin{center}
\subfloat[]{\includegraphics[width=0.5\textwidth]{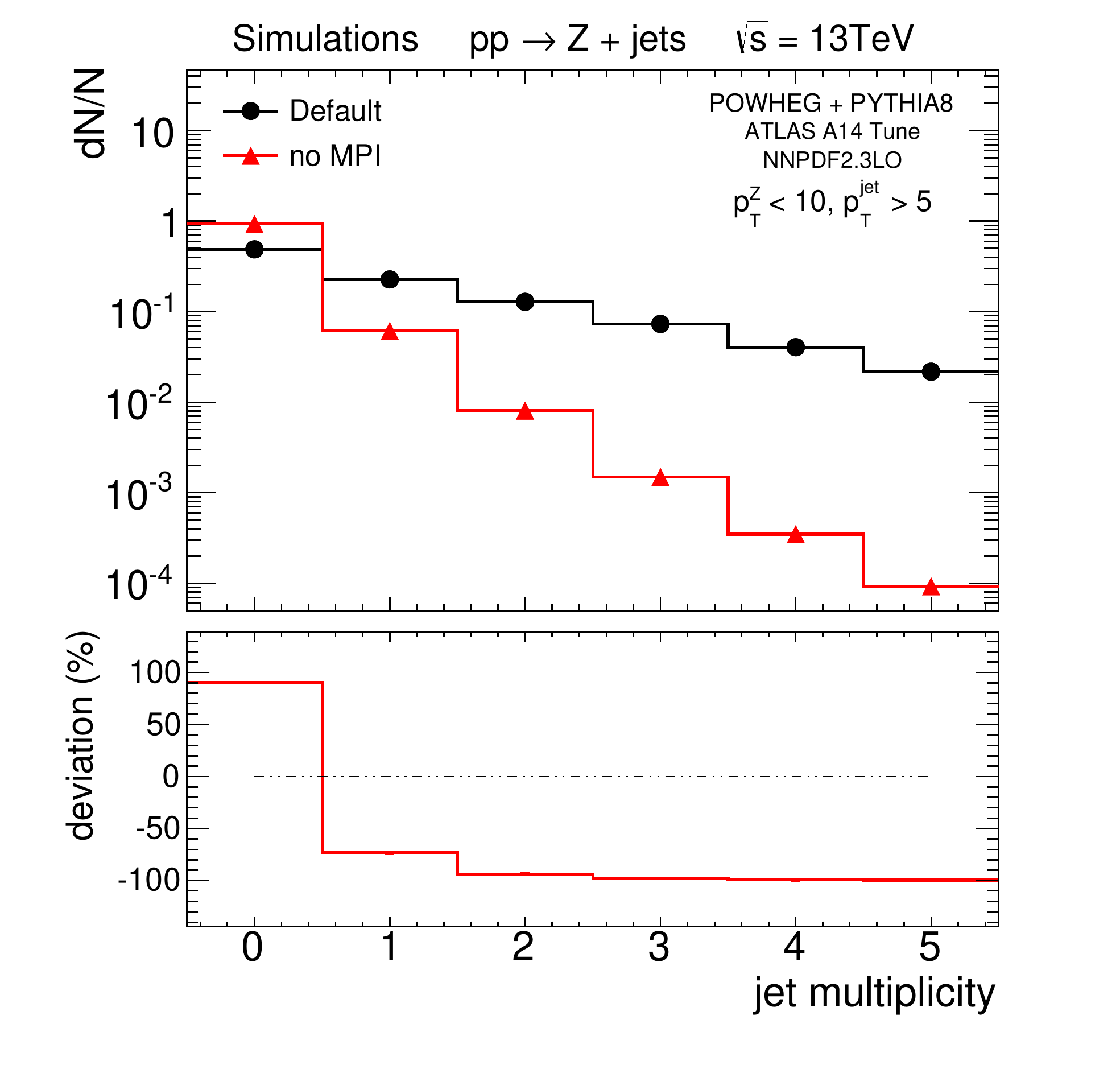}}
\subfloat[]{\includegraphics[width=0.5\textwidth]{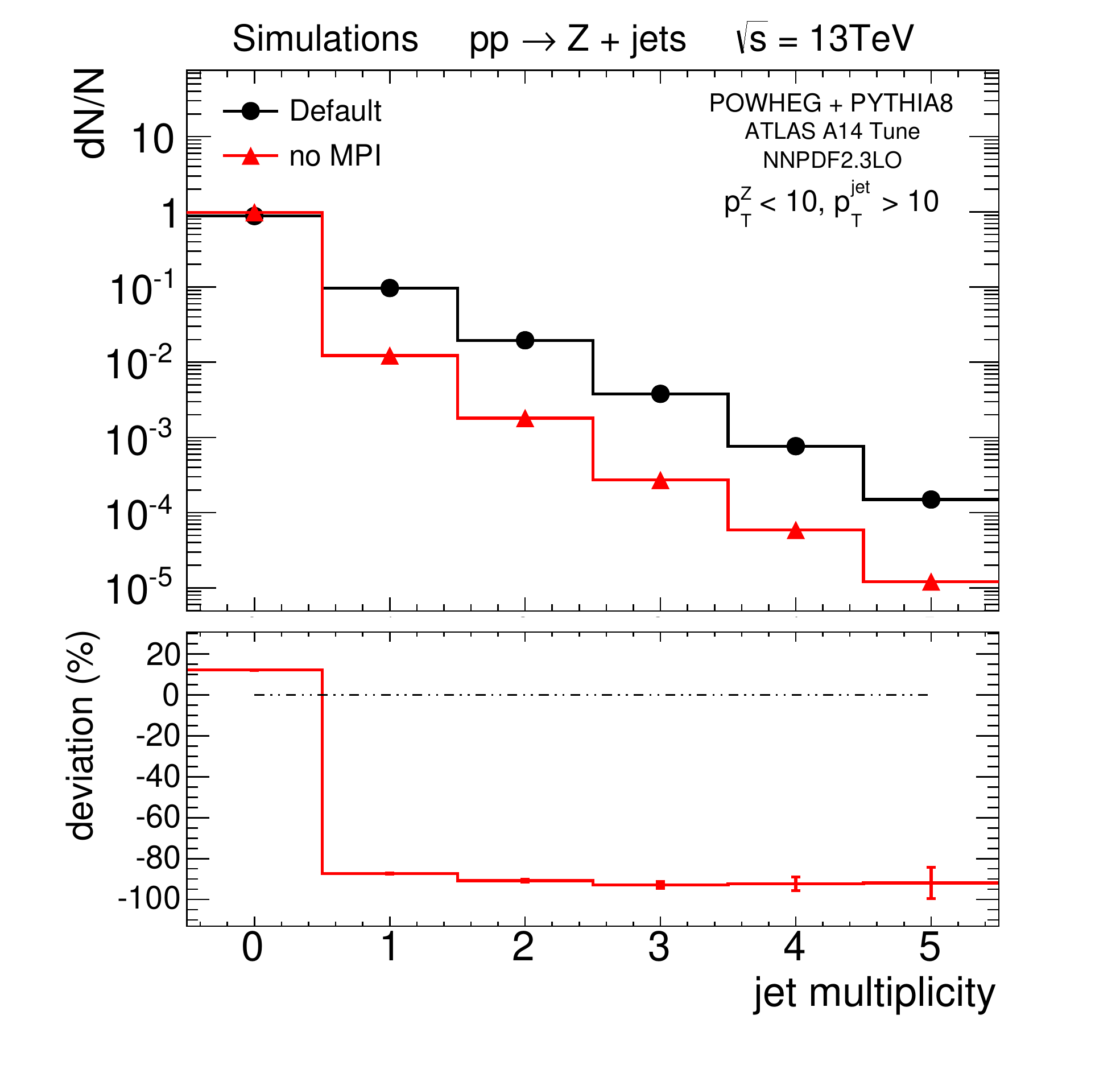}}\\
\subfloat[]{\includegraphics[width=0.5\textwidth]{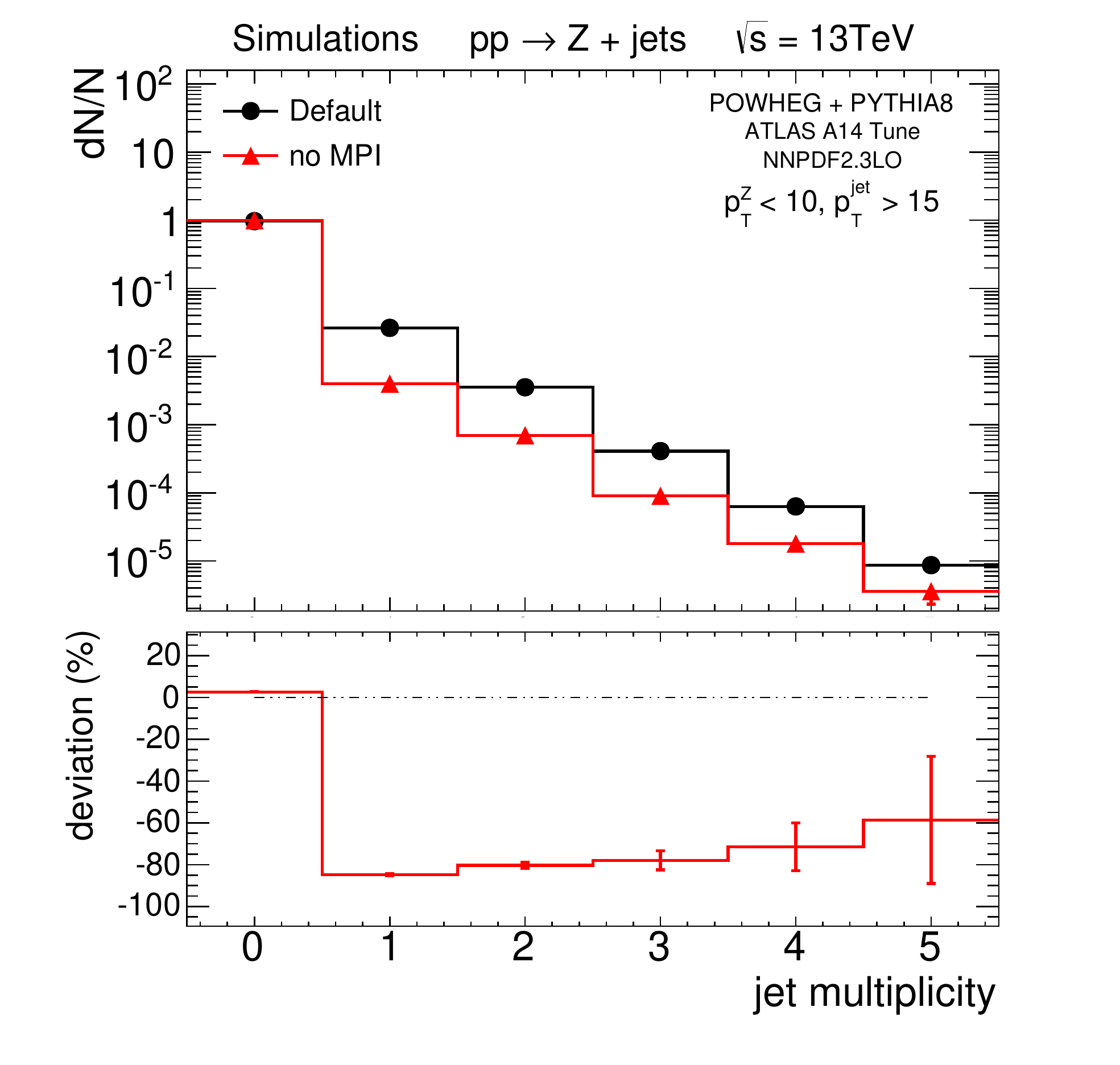}}

\caption {{Jet multiplicity distributions are compared for events with and without MPI. The events from Z + jets 
processes are generated using \textsc{powheg}5, parton showered and hadronized with \textsc{pythia}8.  Jets with 
$p_{\rm T}$ larger than (a) 5 GeV/$c$, (b) 10 GeV/$c$ and (c) 15 GeV/$c$ are  considered and 
$p_{\rm T}^{\rm Z}$ is required to be less than 10 GeV/$c$. The ratio plot in the bottom panel shows 
deviations of the distributions after switching off MPI.}} \label{jm_jetPt}
\end{center}
\end{figure}

Jets with lower $p_{\rm T}$ threshold are selected to study the effect of scale of the second interaction.
Figure~\ref{jm_jetPt}(a), Fig.~\ref{jm_jetPt}(b) and Fig.~\ref{jm_jetPt}(c) shows the sensitivity of jet multiplicity 
with different $p_{\rm T}$ threshold of 5 GeV/$c$, 10 GeV/$c$ and 15 GeV/$c$ of jets respectively. It is clear that MPI 
sensitivity is present even for low $p_{\rm T}$ jets, and hence jet multiplicity can also be used to study the effect 
of energy scale of MPI.

\begin{table}
\caption{\label{tab_varParam}Range of MPI parameters with ATLAS A14 tune and NNPDF2.3LO}
\begin{ruledtabular}
\begin{tabular}{cccc}
Parameter & \textsc{pythia}8 option & default value & variation \\
\hline
${{p_{\rm T}}_{\rm 0}} (\sqrt{s_{\rm 0}})$ [GeV/$c$] & \texttt{MultipartonInteractions:pT0Ref} & 2.09 & $\pm$25\% \\
Z~[cm$^{\rm -1}$] & \texttt{MultipartonInteractions:expPow} & 1.85 & $\pm$25\%\\
$\alpha_{\rm S}^{\rm MPI}$ & \texttt{MultipartonInteractions:alphaSvalue} & 0.126 & $\pm$25\% \\
\end{tabular}
\end{ruledtabular}
\end{table}

\newpage

\section{Summary}\label{sec:summary}
This paper presents the analysis of the Z + jets events to explore the new observables and phase-space region which can 
enhance the sensitivity to the MPI. The Z + jets events are generated with \textsc{powheg}, followed by
 the hadronization and showering with \textsc{pythia}8. The distributions of jet multiplicity associated with Z-boson 
can be useful in inclusive study of MPI.
The sensitivity to the presence of MPI increases significantly by requiring an upper cut on the $p_{\rm T}$ of Z-boson.
It is observed that parameters of the MPI model, have increased sensitivity in the jet multiplicity distribution than 
the correlation observables.
Hence jet multiplicity distribution associated with Z-boson can be used to perform the inclusive MPI measurements 
at the LHC and constraint MPI model parameters with better precision.

\begin{acknowledgments}
This work is financially supported by Department of Science and Technology (DST), New Delhi and University Grant Commission (UGC), New Delhi.
Authors would like to thank Hannes Jung, Paolo Bartalini and Paolo Gunnelini for the providing feedback on the topic.
Authors would also like to thank authors of \textsc{powheg} and \textsc{pythia}8.
\end{acknowledgments}

\newpage

\bibliography{paper_Zjets_PRD.bib}

\end{document}